\documentclass[graybox]{svmult}

\usepackage{mathptmx}       
\usepackage{helvet}         
\usepackage{courier}        
\usepackage{type1cm}        
\usepackage{makeidx}         
\usepackage{graphicx}        
\usepackage{multicol}        
\usepackage[bottom]{footmisc}

\makeindex             

\begin{document}

\title*{Discovery of Peculiar Double-Mode Pulsations and Period Doubling in
        {\it Kepler} RRc Variables}
\titlerunning{Peculiar Double-Mode Pulsations and Period Doubling in {\it Kepler} RRc Stars}
\author{P. Moskalik, R. Smolec, K. Kolenberg, J. Nemec, A. Kunder, M. Chadid,
        G.\thinspace Kopacki, R.~Szab\'o + WG\#13 members}
\authorrunning{Moskalik et al.}
\institute{P. Moskalik  \at Copernicus Astronomical Center, Warszawa, Poland,
                            \email{pam@camk.edu.pl}
\and
           R. Smolec    \at Institut f\"ur Astronomie, Universit\"at Wien, Austria
\and
           K. Kolenberg \at Harvard College Observatory, Cambridge, MA, USA \&
                            Instituut voor Sterrenkunde, K.U. Leuven, Belgium
\and
           J. Nemec     \at Department of Physics and Astronomy, Camosun College,
                            Victoria, British Columbia \&
                            International Statistics and Research
                            Corporation, Brentwood Bay, British
                            Columbia, Canada
\and
           A. Kunder    \at Cerro Tololo Inter-American Observatory, La Serena, Chile
\and
           M. Chadid    \at Observatoire de la C\^ote dAzur, Universit\'e Nice Sophia-Antipolis, France
\and
           G. Kopacki   \at Astronomical Institute, University of Wroc{\l}aw, Wroc{\l}aw, Poland
\and
           R. Szab\'o   \at Konkoly Observatory, Budapest, Hungary}
%
%
\maketitle

\abstract*{We analyzed photometry of the first overtone
RR~Lyrae-type stars (RRc stars) observed by the {\it Kepler}
telescope. All studied RRc variables turned out to be double-mode
pulsators, with a puzzling period ratio of $P_2/P_1 = 0.612-0.632$.
We detected subharmonics of the secondary frequency at $\sim 1/2
f_2$ and $\sim 3/2 f_2$. Their presence is a signature of a
period-doubling of the secondary oscillation.}


We analyzed the Long Cadence (29.4\thinspace min) photometry of four
first overtone RR~Lyr-type stars (RRc stars) observed by the {\it
Kepler} telescope. Only Q0, Q1 and Q2 datasets were used, spanning
138\thinspace days. Analysis including the remaining {\it Kepler}
data will be discussed elsewhere (Moskalik et al. 2012, in
preparation).

\begin{figure}[t]
\sidecaption[t]
\includegraphics[scale=.3]{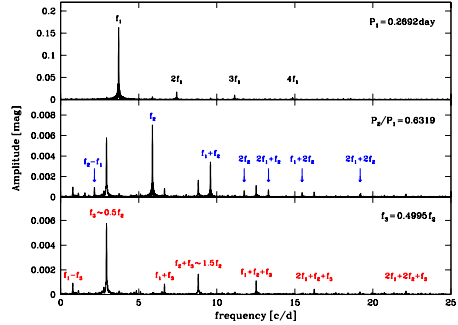}
\caption{Prewhitening sequence for KIC\thinspace 5520878. Upper
panel: FT of original data. Middle and bottom panels: FT after
consecutive prewhitening steps.}
\label{fig1}
\end{figure}

The frequency analysis was conducted with the standard consecutive
prewhitening technique (see e.g. \cite{MK}). All four RRc variables
turned out to be multi\-periodic. Fig.\thinspace \ref{fig1} shows
the prewhitening sequence for KIC\thinspace 5520878. After removing
the do\-minant frequency $f_1 = 3.715$\thinspace c/d and its
harmonics, the Fourier transform of the residuals (middle panel) is
dominated by a strong peak at $f_2 = 5.879$\thinspace c/d, its
harmonic and their combinations with the primary frequency, $f_1$.
The period ratio of the two modes is $P_2/P_1 = 0.6319$. After
prewhitening the data with both $f_1$, $f_2$ and their harmonics and
combinations (bottom panel), the strongest remaining signal appears
at $f_3=2.937$\thinspace c/d, {\it i.e} at $\sim 1/2 f_2$. Thus,
$f_3$ is not an independent frequency, but a {\it subharmonic} of
$f_2$. The second subharmonic at $\sim 3/2 f_2$ is also clearly
visible. All the remaining peaks in the Fourier transform (FT)
correspond to combination frequencies.

The frequency spectra of the other three {\it Kepler} RRc stars are
very similar. The strongest secondary peak always appears at
$f_2/f_1 = 1.58-1.63$ (or $P_2/P_1 = 0.612-0.632$). Also, in each
star we detect at least one subharmonic of $f_2$, either at $\sim
1/2 f_2$ or at $\sim 3/2 f_2$. The presence of subharmonics is a
characteristic signature of a {\it period doubling}. After RRab
Blazhko stars \cite{RRPD} and BL~Her-type stars \cite{Smolec}, the
RRc variables are the third class of pulsators in which period
doubling has recently been found.

Peculiar period ratios of $\sim 0.60-0.63$ are not unique to the RRc
stars observed by {\it Kepler}. They have also been found in 9 other
RR~Lyr-type variables \cite{AQLeo,OmCen,LMCRR}. Except for the
double mode star AQ~Leo, all the variables belong to the RRc type.
Together with the four {\it Kepler} stars, they constitute a new
class of double-mode RR~Lyr-type pulsators.

Secondary modes with the same puzzling period ratios have also been
detected in classical Cepheids \cite{MK,LMCCep,SMCCep}. Again, they
are found only in the overtone pulsators. Both for the Cepheids and
for the RR~Lyr-type stars, the secondary frequency, $f_2$, falls
{\it between} the frequencies of the third and the fourth radial
overtones \cite{DS}. This implies that $f_2$ must correspond to a
{\it nonradial} mode of oscillation.

\end{document}